\documentclass[11pt,twoside]{article}
\usepackage{asp2014}

\aspSuppressVolSlug
\resetcounters

\begin{document}
\title{Operations in the era of large distributed telescopes}

\author{Yan Guillaume Grange$^1$, Kevin Vinsen$^2$, Juan Carlos Guzman$^3$, Jos\'{e} Alfredo Parra$^4$, Jan David Mol$^1$, Rosly Renil$^5$, and Christoper Schollar$^5$}
\affil{$^1$ ASTRON, P. O. Box 2, 7990AA Dwingeloo, The Netherlands \email{grange@astron.nl}}
\affil{$^2$ ICRAR, Perth, Western Australia}
\affil{$^3$ CSIRO Astronomy and Space Science, Australia}
\affil{$^4$ ALMA Observatory, Chile}
\affil{$^5$ SKA SA, MeerKAT, South Africa}   
    
\paperauthor{Yan Grange}{grange@astron.nl}{0000-0001-5125-9539}{ASTRON}{}{City}{Dwingeloo}{7990AA}{The Netherlands}
\paperauthor{Kevin Vinsen}{kevin.vinsen@uwa.edu.au}{0000-0001-5332-3784}{University of Western Australia}{ICRAR}{Perth}{WA}{6009}{Australia}
\paperauthor{Juan Carlos Guzman}{juan.guzman@csiro.au}{}{CSIRO Astronomy and Space Science}{}{Perth}{WA}{}{Australia}
\paperauthor{Jos\'{e} Alfredo Parra}{Jose.Parra@alma.cl}{}{ALMA Observatory}{}{Santiago}{}{}{Chile}
\paperauthor{Jan David Mol}{mol@astron.nl}{}{Author1 ASTRON}{}{City}{Dwingeloo}{7990AA}{The Netherlands}
\paperauthor{Rosly Renil}{rosly@ska.ac.za}{}{SKA South Africa}{}{Cape Town}{}{}{South Africa}
\paperauthor{Christopher Schollar}{rosly@ska.ac.za}{}{SKA South Africa}{}{Cape Town}{}{}{South Africa}

\begin{abstract}
The previous generation of astronomical instruments tended to consist of single receivers in the focal point of one or more physical reflectors. 
Because of this, most astronomical data sets were small enough that the raw data could easily be downloaded and processed on a single machine. 

In the last decade, several large, complex Radio Astronomy instruments have been built and the SKA is currently being designed. 
Many of these instruments have been designed by international teams, and, in the case of LOFAR span an area larger than a single country. 
Such systems are ICT telescopes and consist mainly of complex software. 
This causes the main operational issues to be related to the ICT systems and not the telescope hardware. 
However, it is important that the operations of the ICT systems are coordinated with the traditional operational work. 
Managing the operations of such telescopes therefore requires an approach that significantly differs from classical telescope operations.

The goal of this session is to bring together members of operational teams responsible for such large-scale ICT telescopes. 
This gathering will be used to exchange experiences and knowledge between those teams. 
Also, we consider such a meeting as very valuable input for future instrumentation, especially the SKA and its regional centres.
\end{abstract}
\section{Introduction}
The authors formed a panel at ADASS XXVI in Trieste, Italy on the evening of 17$^th$ October 2016 for a Birds of a Feather Session to look at ``Operations in the era of large distributed telescopes''. 
The suggested items to ve consider were as follows:
\begin{itemize}
\item Experiences in the area of transitioning from Construction to Operations
\item Integration of HPC environments (usually located off-site) into telescope operations and scheduling.
\item Need for service level agreements
\item Monitoring of system parts out of our control (like networking)
\item Culture (in international teams)
\item Scaling up to order of magnitudes larger number of observations
\item RFI (which also scales with size)
\item There is no driver for an archive because funding comes from people who write proposals. For those, the archive is only a way to provide data to the PI. 
\item Acquiring funding for (pure) operations causing the use of R\&D grants to acquire operational hardware or development, which is risky business.
\end{itemize}

The aim was not to solve all these problems in the one session, but to provoke discussion.
This paper gives a brief overview of the salient points of the discussion.

\section{Transition from Construction to Operations}

\begin{itemize}
\item \textbf{Monitoring of system parts out of our control.}\\
Monitoring of parts that are out of the control of the operations teams is very hard to solve. 
The operations team need a good working relationship with the people on the ground; the partners of the telescope. 
Sometimes the operations teams notice more errors than the other partners own monitoring and you have to build up a trust relation. 
This is hard to build in to Service Level Agreements (SLA).

\item \textbf{Use case for commissioning aren't the same for operations.}\\
The use cases for commissioning are very different from the system that was designed.
This is particular true for the automatic pipelines. 

In most cases the early science is not generated from an automatic pipeline.
It is individuals running the various stages by hand.

\item \textbf{Knowledge Transfer.}\\
Transferring all the expertise in the engineering team to the operational team is very important, since it reduces possible down time. 
That transfer can take time. 
Since the time line to get the instrument running is generally challenging, this process needs to be in place.
When moving the telescope from development to operations, and doing early science, you are in a phase where you have an operational telescope while you are still changing the system. 
You want to minimise down time.

\end{itemize}

\section{High Performance Computing Centres}

\begin{itemize}
\item \textbf{Integration of HPC environments into telescope operations and scheduling.}\\
HPC systems have to be integrated into the instrument. 
It doesn't stop at the correlator. 
If the supercomputer goes down, you have to stop observing. 
The people who work in HPC centres have a different culture. 
They are used to work being submitted as batch jobs. If a system goes down, it is not a  major issue as the jobs will restart.
If data is being streamed into the HPC this is a major issue.
The mandate of research HPC is to provide computing. 
However the type of usage for Radio Astronomy is very different to what they are used to. 
So the research HPC environments have to start thinking how to design their systems and operations to have redundancy so that you can have continuity of service and uptime. 

\item \textbf{Cloud Computing.}\\
Using cloud computing, one can not really trust they are future proof. 
Storing Terabytes of data has major cost implications.
Silly and simple mistakes in scripts can give, quite unintentional, very high bills. 
However experimenting with using the cloud gives the HPC centres a sense of competition and may actually give them an working examples of good SLAs that work in industry.

\item \textbf{Need for strong service level agreements.}\\
You need the HPC facility running when data is arriving from the telescope.
The HPC facility needs to work towards this which means the mandate must come from 
Written service agreements signed by high-level people.

\item \textbf{Issues with organisations to store data.}\\
Many of the organisations storing data for projects use infrastructure that is Government funded.
If the Government reduces or removes the funding the data may well be lost.
Unfortunately long term archives have been an afterthought to most, but not all, major telescopes in the past decade. 

\item \textbf{Framework for running the pipelines.}\\
Most astronomy work is not well suited to batch jobs.
Currently the most popular way of processing astronomy data is to define workflow components statically in scripts.
These scripts are then either executed sequentially on a local machine or wrapped into job scripts submitted to job scheduling systems in an HPC environment.
Such application-driven workflow models have several drawbacks.
Firstly, maintainable; secondly, the execution process lacks real-time monitoring and control; and finally, even if failures or exceptions are noticed at an earlier stage users still have to restart the entire job for re-execution. 

\end{itemize}

\section{Team Interaction}

\begin{itemize}
\item \textbf{Culture (in international teams).}\\
Video and Conference calls only go so far - face to face meetings are still necessary.
Face to face interaction is very important to understand each others needs and possibilities. 
This is not just for the Scientists but includes the HPC support people. \\
You may expect that other countries work the same way yours do. 
This is not the case.
This can result in asking questions to the wrong people in the hierarchy for example, and delay a project. 

\item \textbf{Commercial companies don't understand science requirements. Cultural difference between what we and what they measure.}\\
For the Astronomy team it can be really annoying if we have 0.01\% data loss.
To a commercial company this might be perfectly acceptable.
The telescope records data in real-time, and with data loss you get holes which grow larger after processing. 
\end{itemize}

\section{Code Development and Release}

\begin{itemize}
\item \textbf{Don't worry too much about the end product; concentrate of what's next.}\\
Many systems have spent to much time in analysing requirements, for the full system and have forgotten that there is a long road to go for the full system, and that the transition from construction to operation doesn't follow a waterfall model.
This is compounded by the face Along the way, requirements have to change and resources have to be diverted to try and help the project going forward.

\item \textbf{Coders having different objectives.}\\
One of the main challenger has been to make the various groups of software developers to work together. 
They share different objectives (developers want to create code as fast as possible, while operations want to have a stable system). 
They normally do not know each others requirements and constraints, especially when they are geographically distributed. 

\item \textbf{Maintenance} \\
One key issue in software management is software maintenance. 
One specific issue is software obsolescence. 
Some telescopes have been running for over 50 years. 
The main problem is that you can't predict the funding of your telescope on the long run you don't know whether investing in new software is worth it for a longer time. 
A pragmatic approach is that if support stops, change or upgrade software. 
So you have to make sure software keeps being upgradable.

\end{itemize}

\section{The Archive}

\begin{itemize}
\item \textbf{There is no driver for an archive because funding comes from people who write proposals. For those, the archive is only a way to provide data to the PI.}\\
MeerKAT integrated the archive as part of the commissioning phase. 
They had the PIs interact with the archive staff and get involved in meta data extraction. 

LOFAR didn't suffer from an outdated design of the archive since the archive wasn't designed in from the start. 
The focus was mostly on getting the instrument built. 
The archive should have been part of the commissioning of the instrument. 
This is especially important because when you see that for some telescopes, about half the scientific output comes from the archive. 

In ALMA, the data is now distributed to the regional centres. 
The commissioners have access to the local repository in Chile. 
This may have been an issue in the beginning. 
ALMA was very modern in the sense that the archive was designed as a central part from the beginning. 
However it is generally underestimated how much work data management was, and ALMA was not an exception to that. 
The good news is that this seems to be developing. 
LSST is for example spending about half the budget to data management.

\end{itemize}

\end{document}